\documentclass[usenatbib]{mn2e}

\usepackage[dvips]{graphicx}
\usepackage{amsmath,amsfonts,amssymb}
\usepackage{times} 
\usepackage{color}

\usepackage{hyperref}
\hypersetup{colorlinks=true,
            citecolor=blue,
            linkcolor=blue}

\definecolor{orange}{rgb}{1,0.5,0}

\def\mi{m_i}
\def\mt{m}
\def\mui{\mu_i}
\def\mut{\mu}
\def\ai{a_i}
\def\at{a}
\def\ei{e_i}
\def\et{e}
\def\qi{q_i}
\def\qt{q}
\def\Qi{Q_i}
\def\Qt{Q}
\def\di{\delta_i}
\def\Frac#1#2{{{\displaystyle\strut#1}\over{\displaystyle\strut#2}}}

\begin{document}

\title[A semi-empirical stability criterion for real planetary systems]{A semi-empirical stability criterion for real planetary systems with eccentric orbits}
\author[Giuppone et al.]{C. A. Giuppone$^{1,2}$, M. H. M. Morais$^{1}$, A. C. M. Correia$^{1,3}$\\
$^{1}$ Departamento de F\'isica, I3N, Universidade de Aveiro, Campus de Santiago, 3810-193 Aveiro, Portugal\\
$^{2}$ Observatorio Astron\'omico, IATE, Universidad Nacional de C\'ordoba, Laprida 854, 
(X5000BGR) C\'ordoba, Argentina (cristian@oac.uncor.edu)\\
$^{3}$ ASD, IMCCE-CNRS UMR8028, Observatoire de Paris, UPMC,
77 Av. Denfert-Rochereau, 75014 Paris, France  \\
}

\date{}

\maketitle

\begin{abstract}
We test  a crossing orbit stability criterion for eccentric planetary systems, based on Wisdom's criterion of first order mean motion resonance overlap \citep{Wisdom1980AJ}. 
We show that this criterion fits the stability regions in real exoplanet systems quite well. 
In addition, we show that elliptical orbits can remain stable even for regions where the apocenter distance of the inner orbit is larger than the pericenter distance of the outer orbit, as long as the initial orbits are aligned.
The  analytical expressions provided here can be used to put rapid constraints on the stability zones of multi-planetary systems. 
As a byproduct of this research, we further show that the amplitude variations of the eccentricity can be used as a fast-computing stability indicator.
\end{abstract}

\section{Introduction}
Multi-planetary systems are discovered and confirmed with increasing frequency, with many of the newest planets in configurations called ``compact systems'' (where all possible stable regions are occupied and no additional bodies can be found). 
Usually the discoveries of new planetary systems are part of large projects, and it is important to know which systems are already ``full'' or not, in order to look for additional companions. In addition, proposals for observations focus in discovering Earth-like planets in already confirmed exo-systems. It is thus important to have a fast criterion to evaluate the stability  of Earth-like planets in  already confirmed exo-systems.

The Circular Restricted 3-Body Problem (CR3BP) consists of a particle moving  under the gravitational influence of a primary and secondary masses. The primary and secondary move in a circular orbit about their common centre of mass and the particle is too small to affect the other two. There is a conserved quantity called \textit{Jacobi constant}, that can be used to determine regions of allowed motion. The region of gravitational influence of the secondary mass is bounded by the Lagrange points, $L_1$ and $L_2$, forming the Hill sphere. The  \textit{Jacobi constant} can be interpreted as the ``energy'' of the test particle and defines the region of allowed motion.  A particle can remain confined in orbit around the primary, or it can cross the Hill sphere through $L_1$, or even escape through $L_2$. A particle that cannot cross the Hill-sphere is called \textit{Hill-stable} \citep[see e.g.][]{Murray_Dermott_1999}.

In the late 19th century Henri Poincar\'e  studied the stability of the  three-body problem. He hinted at the complicated nature of the motion that can arise for some starting conditions.  Some trajectories, called chaotic, diffuse in phase-space and { may escape or collide with one of the bodies}, in contrast with regular trajectories \citep{Poincare_1899}. { In the CR3BP,  orbits that do not obey Hill-stability but are regular remain bounded, while chaotic orbits that  do not obey Hill-stability may become unbounded}.  

\citet{Wisdom1980AJ} deduced a criterion for the onset of chaos in the CR3BP based on the overlap of first order mean motion resonances. The overlap extends in a region around the planet of width 
\begin{equation}
\delta =C \mu^{2/7} a \label{eq.wis},
\end{equation}
where $\mu$ is the mass ratio between the planet and its parent star,  $a$ is the semi-major axis of the planet, and $C$ is a constant value.
\citet{Wisdom1980AJ} obtained a theoretical value $C=1.33$, but using numerical simulations
\citet{Duncan_etal_1989} estimated $C=1.57$. The orbits of test particles in this region exhibit chaotic diffusion of eccentricity and semi-major axis until escape or collision occurs. 

\citet{Marchal&Bozis} obtained Hill-stability criteria in the general 3-body problem. These apply, in particular, to  systems with two-planets orbiting a central star. In this case, the stability limit in terms of total angular momentum, $c$, and total  energy, $h$, is:
\begin{equation}\label{eq.12.marchal}
 - \frac{2 M}{ M_{\star}^3 } \frac{c^2 h}{{\cal G}^2} > 1 + 3^{4/3}  \frac{\mu_1 \mu_2}{\alpha^{4/3}} \ , 
\end{equation}
where $\cal G$ is the gravitational constant, $\mi$ is the planetary mass,  $m_{\star}$ is the stellar mass, $M=m_\star + m_1 + m_2$, $M_\star=m_{\star}m_1+m_{\star}m_2+m_2 m_1$, $\mui=\mi/m_\star$, and $\alpha=\mu_1+\mu_2$.

\citet{Gladman_1993} studied the stability of two close planets numerically and compared results with existent analytic criteria.  He applied \citet{Marchal&Bozis} criterion to provide a  relation in function of orbital elements. This was obtained by rewriting expression (\ref{eq.12.marchal}) as:
\begin{equation}\label{eq.21.gladman}
 \left( \mu_1+ \mu_2 \frac{a_1}{a_2} \right) \left( \mu_1 \gamma_1 + \mu_2 \gamma_2 \sqrt{\frac{a_2}{a_1}} \right)^2 >  \alpha^{3} + 3^{4/3}  \mu_1 \mu_2 \alpha^{5/3}, 
\end{equation}
where $\gamma_{i} = \sqrt{1 - \ei^2}$,  $\ei$ are the planets' eccentricities, and $\ai$ are the planets' semi-major axis.
Note that if $\mu_1$ or $\mu_2$ are zero (restricted problem) the above relation becomes $\gamma_i^2 > 1$, that is, $\ei < 0$ in the non-circular cases, so Hill stability cannot be obtained directly \citep{Marchal&Bozis}. 
Expressions (\ref{eq.12.marchal}) and (\ref{eq.21.gladman}) are then better suited to evaluate systems of comparable planetary masses.

Writing $a_2=a_1 (1+\delta)$, we can solve Eq.~(\ref{eq.21.gladman}) to obtain an estimate of the Hill stability region for the planet $m_2$  as function of $e_2$,  by fixing  $a_1$ and $e_1$ for the planet $m_1$. Moreover, when the planets have circular orbits, the previous expression simplifies as \citep[Eq.\,24,][]{Gladman_1993}:
\begin{equation}
\delta>3.46 \, \left[ \frac{\mu_1+\mu_2}{3} \right]^\frac{1}{3} a_1 \label{eq.del} \ ,
\end{equation}

In many works on stability, $\delta$ is often defined in terms of mutual Hill radius $R_H$:
\begin{equation}
 R_H = \left[\dfrac{\mu_1+\mu_2}{3}\right]^{1/3} \frac{a_1+a_2}{2}.
\end{equation}
Hence we can write Eq. (\ref{eq.del}) as
\begin{equation}
\delta>3.46 \, R_H \label{eq.gl},
\end{equation}

The stability of coplanar low eccentricity equally-spaced multiple planet systems has  been investigated numerically by several authors. 
\citet{Chambers_1996} showed that in the  mass range $10^{-6} -10^{-2}\,M_J$  the mutual distances should be at least $7-9$ mutual Hill radii to ensure stability on a 10-Gyr timescale. 
\citet{Smit&Lissauer} performed simulations for 10-Gyr of  multi-planetary systems of $1 \, M_{\oplus}$ orbiting a Sun-like star and also considering a Jupiter-like planet as perturber in some simulations. According to their results, in closely packed systems  stability for more than $10^7 - 10^8$ years is assured when the mutual distances are larger than $10\,R_H$. 
However, these numbers should not be taken too exactly because they were obtained assuming regularly-spaced, equal-mass planetary systems. They are also not applicable
to eccentric orbits and dynamical configurations such
as mean-motion resonances. 
For instance, in the GJ\,876 planetary system the separation between the two Jupiter-size planets in the 2/1 mean motion resonance is only one Hill radius \citep{Correia_etal_2010}.
Moreover, \citet{Lovis_etal_2011} (Fig.13) show that several known multi-planetary systems
are dynamically ``packed'', with mutual separations less than $10\,R_H$. 

{ \citet{Barnes_2006b} performed 1000 numerical integrations of planetary systems similar to HD\,12661 and 47\,UMA, which have Jupiter-mass planets and moderately eccentric orbits. 
Their work suggests that  the true instability boundary is roughly 10\% greater than the Hill-stability boundary of \citet{Marchal&Bozis} 
(Eq.\ref{eq.21.gladman}).} { This confirms that  although Hill-stability does not guarantee Lagrange-stability, the two boundaries may be close, at least for planet systems with moderately eccentric orbits.  A planetary system near the Hill-stability boundary may become unstable if the inner planet collides with the star, or if the outer planet escapes, although collisions between the two planets cannot occur.}

\citet{Quillen2011} computed the strengths of zero-th order (in eccentricities) three-body resonances for a co-planar  multi-planetary system. Their analytical estimates, assuming equal mutual distances and equal planet to star mass ratio $\mu$ , show that these resonances overlap when
\begin{equation}
\delta\leq 2 \, \mu^{1/4} a \ , \label{Quillen11}
\end{equation}
For mass ratios $\mu\sim 10^{-3}-10^{-6}$ this limit is about 2 times Wisdom's 2-body resonance overlap limit defined in Eq. (\ref{eq.wis}).  
 
We just saw that analytic stability criteria were obtained in very specific cases, namely the CR3BP, 2-planet systems and 3-planet systems with circular orbits. Moreover, while Hill-type criteria guarantee stability in the three-body problem, there are regular orbits outside Hill-stable regions that do not escape. An accurate stability criterion must rely on obtaining the regions where large scale chaotic diffusion is possible but this has only been done in the case of circular 2-planet and 3-planet systems. Finally, current empirical stability criteria for multi-planet systems rely heavily on numerical integrations, thus may be affected by the choice of initial conditions. In particular, we observe that these empirical criteria often underestimate the size of the stability regions.

Our aim is to obtain a stability criteria with the purpose of providing a quick guide to exo-planet discovery teams. We considered non-equally-spaced real planetary systems with mass ratio ranging from 2 to 100, covering Earth-like and Jupiter-like planets. We provide some guidelines so that observers can quickly  test the stability without having to rely on  time-consuming numerical integrations.

\section{Crossing orbits criterion}
\label{coc}

A crossing orbit stability criterion relies on the notion that if two orbits intercept at some point,  unless there is some kind of resonant mechanism, close encounters occur and the system becomes destabilized. Therefore, we expect that the  stability limits should approximately follow the pericentric and apocentric collision lines. We will explain this concept better below.

Consider a system composed by two planets. One planet, with mass $\mi$, is a real planet whose orbital parameters $\ai$ and $\ei$ are known. The other planet, with mass $\mt$, is a ``test'' planet, whose orbital parameters ($\at$, $\et$) are unknown.
In stability studies, we usually want to know which sets of parameters ($\at$, $\et$) are stable or not.

\begin{figure}
 \begin{center}
   \begin{tabular}{c}
\includegraphics*[width=5.cm]{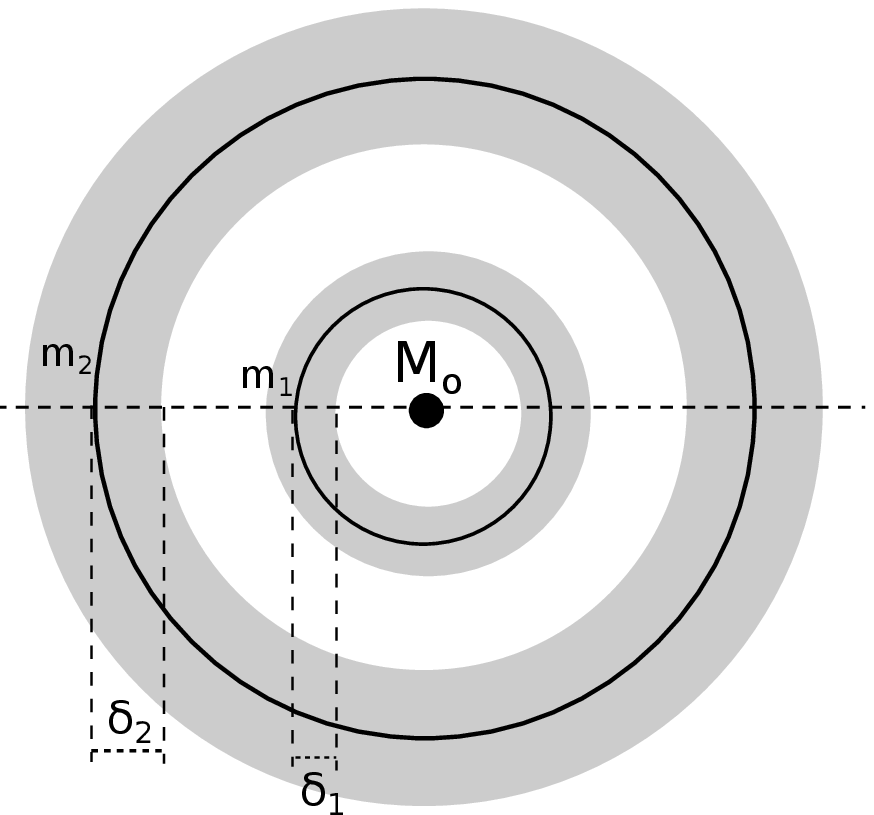} \\                                    
\includegraphics*[width=5.cm]{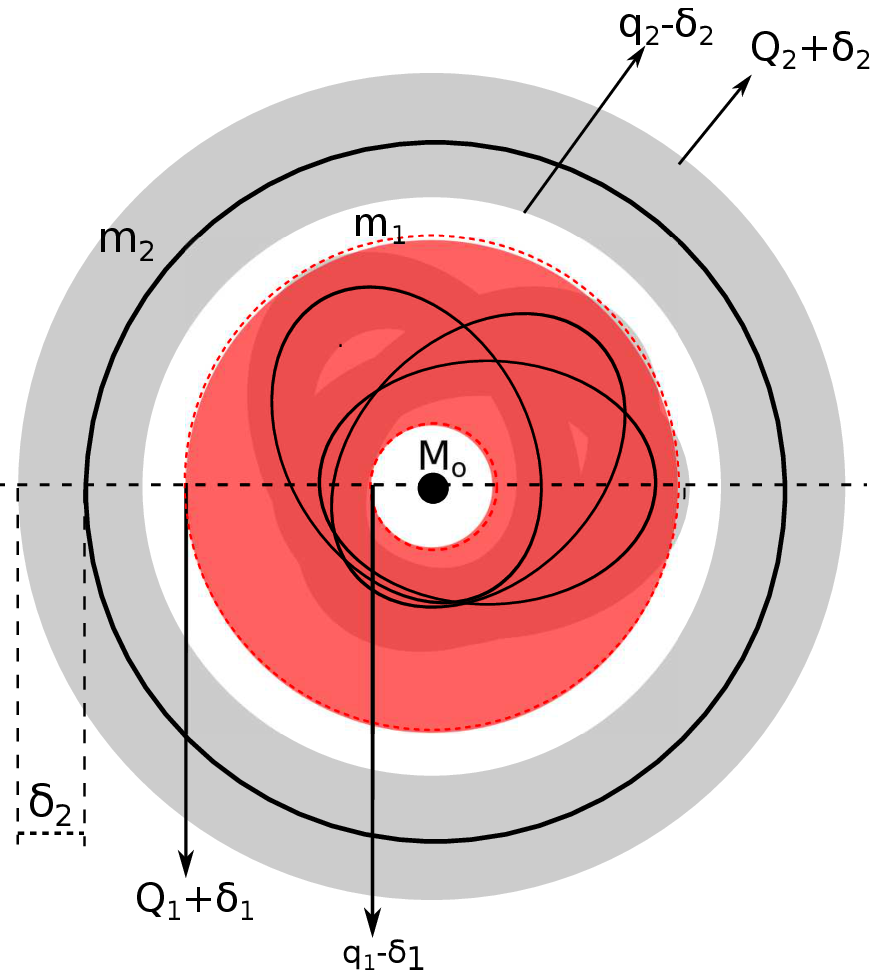}
\end{tabular}
 \caption{Stability regions schema for a system with 2 planets. Top: Both planets are in circular orbits. The unstable region fills a certain space $\ai \pm \di$ (grey zone). Bottom: One planet has an elliptical orbit. The eccentric orbit precesses with time, so the unstable region is delimited by the pericentric $\qi$ and apocentric $\Qi$ distances.} 
\label{fig.stability}
 \end{center}
\end{figure}

Assume that the two planets $\mi$ and $\mt$ have circular orbits. Around each body there exists a region of instability with radius $\di$. 
As time evolves, the position of each planet changes, and after several periods, the instability zone for each planet covers a strip around the orbit (Fig. \ref{fig.stability}). When one of the planets is on a eccentric orbit, the longitude of pericenter circulates and the orbit precesses. Consequently, for different times we have an instantaneous ellipse with an unstable region around it. Since the ellipse is precessing, the instability region is extended.
Therefore, if we considerer eccentric orbits, the stability limits  should approximately follow the pericentric and apocentric collision lines (Fig. \ref{fig.stability}).

\begin{figure}
 \begin{center}
\includegraphics*[width=6.cm]{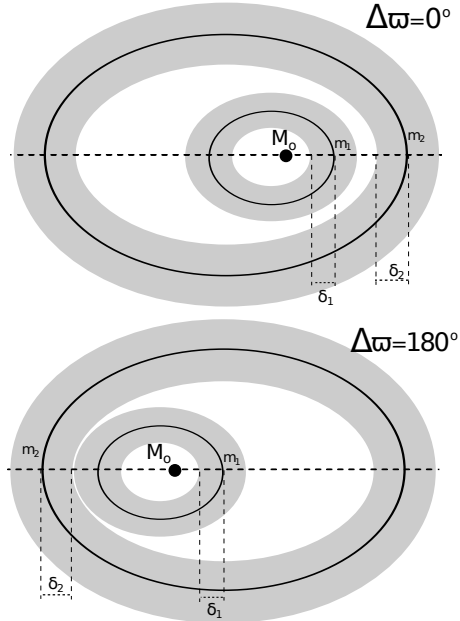}
 \caption{Stability regions schema for a system with 2 planets in eccentric orbits. Top: the orbits are initially aligned. Bottom: the orbits are initially anti-aligned. The first situation allows closer semi-major axis than the second one. If somehow the system is able to keep its initial configuration, stability regions are wider in the first case.} 
\label{fig.stability2}
 \end{center}
\end{figure}

The pericentric $\qt = \at (1-\et) $ and apocentric $\Qt = \at (1+\et)$ distances correspond to the extreme orbital positions, so they give the minimum distances for close encounters with an object in a inner or outer orbit, respectively. In a real interacting system, these values change in time but we are only concerned with the initial observed orbital elements. Moreover, the closest approach depends not only on the pericentric ($\qt$) and apocentric ($\Qt$) distances of both orbits, but also on the orbits' initial alignment, which is given by the difference in the longitudes of the pericenters, $\Delta \varpi $. 
Indeed, if the orbits are initially aligned ($\Delta \varpi = 0^\circ$) the minimum distance between the planes is maximized, while for anti-aligned orbits ($\Delta \varpi = 180^\circ$) it is the contrary (Fig.\ref{fig.stability2}).

\begin{figure}
 \begin{center}
   \begin{tabular}{c}
\includegraphics*[width=8.cm]{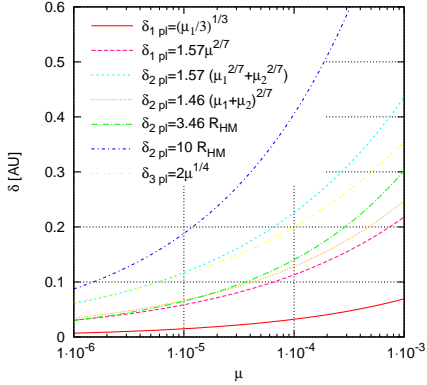}
\end{tabular}
  \caption{ Comparison of $\di$ for the different instability criteria described in this paper as a function of the mass-ratio to the star $\mu$. The planets have equal masses, the central star  has $ 1 M_\odot$, and the inner planet' semi-major axis is $\ai=1$~AU. In red we plot the contribution of two individual Hill-regions. The green line is calculated using the mutual Hill radius as deduced by Gladman for two-planets on circular orbits. In blue, we show the 10 mutual Hill radius estimated from some numerical works. The pink line shows $\di$ using Wisdom {resonance overlap criterion} for a single planet. The light-blue line shows the semi-empirical criterion proposed in this paper (Eq.\,\ref{Wisdom1}) { which in this case is a factor 1.76 larger than the resonance overlap criterion for 2 planets obtained by \citet{Deck_etal_2013} (in orange).} Finally, in brown we show Quillen's 3-body criterion.} 
\label{fig-delta}
 \end{center}
\end{figure}

We can construct a crossing orbit stability criterion using the initial values of the pericentric and apocentric distances that define the collision lines: $\Qt=\Qi$ and $\qt=\qi$ for aligned systems; $\Qt=\qi$ and $\qt=\Qi$ for anti-aligned systems.  From the pericentric and apocentric collision lines, we obtain an extended crossing orbit region by  adding (or subtracting) Wisdom's overlap criterion (Eq.\,\ref{eq.wis}) from previous section, defined as\footnote{{ After submission of this article, a new work by  \citet{Deck_etal_2013} was published where the authors develop and test an analytic criterion for the onset of chaotic motion due to overlap of first order mean motion resonances in the  non-restricted three-body problem, valid for small eccentricities. They conclude that the minimum distance between the planets is $1.46\mu_{t}^{2/7}$ where $\mu_t=(m_i+m)/m_{\star}$.  Hence, the minimum distance obtained with Eq.~\ref{Wisdom1} is larger that the estimate of  \citet{Deck_etal_2013}, at most, by a factor $1.76$ in the case of equal masses. In the next section, we will see that in practice our expression provides a reasonable, more conservative, estimate of the size of the chaotic region.}}:
\begin{equation}
\di \equiv C \mui^{2/7} \ai + C \mut^{2/7} \at \approx 1.57 \left[ \mui^{2/7} + \mut^{2/7} \right] \ai  \ , \label{Wisdom1}
\end{equation}
where we adopt the numerical estimate $C=1.57$ \citep{Duncan_etal_1989}.
Since Wisdom's criterion is established using circular orbits, we have chosen to use the semi-major axis in the previous expression. Our assumption is that Wisdom's criterion still provides a reasonable estimate for the size of the chaotic regions when the orbits are not circular. 
Our choice is also one of the most conservative among all criteria described in previous section (see Fig.~\ref{fig-delta}).
Indeed, only the 10 mutual Hill radius provides a larger value for $\di$ (that gives too large instability regions for large mass-ratios).
In Figure~\ref{fig-delta} we also see that for identical low-mass planets (Earth-like to Neptune-like planets) some criteria become equivalent: the Wisdom's overlap criterion for a single planet (Eq.\,\ref{eq.wis}) is similar to the 3.46 mutual Hill Radius deduced by Gladman for two planets (Eq.\,\ref{eq.gl}); and the empirical criterion given by the above equation (\ref{Wisdom1}) for two planets  is similar to Quillen's three-body resonance overlap (Eq.\,\ref{Quillen11}).

In the case of an initially aligned system, the extended crossing orbit region depends on whether $\et$ is higher or  lower than $\ei$
\begin{center}
\begin{tabular}{c c c}     
{\smallskip}
          Interior limit & Exterior limit & \\
\hline\hline\noalign\\
\normalfont
 $ \at > (\ai - \di) \Frac{1-\ei}{1-\et}$ , & $ \at < (\ai + \di) \Frac{1+\ei}{1+\et} $ , & $\et < \ei$ \\
{\smallskip}\\
 $ \at > (\ai - \di) \Frac{1+\ei}{1+\et}$ , & $ \at <  (\ai + \di) \Frac{1-\ei}{1-\et} $ , & $\et > \ei$ \\
{\smallskip}\\ \hline
\end{tabular}
\end{center}

\begin{figure}
 \begin{center}
   \begin{tabular}{c}
\includegraphics*[width=8.cm]{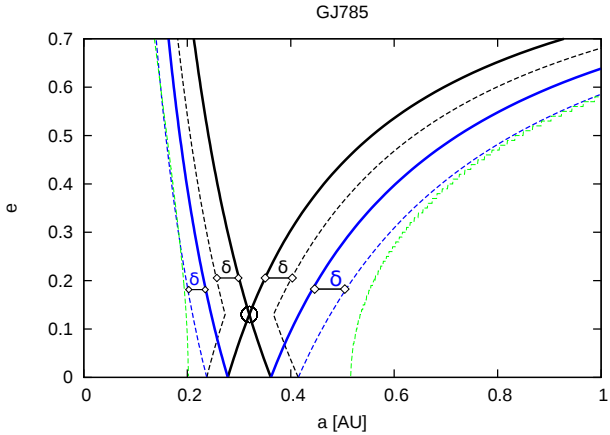}
\end{tabular}
  \caption{Scheme of stability regions in the plane ($\at$, $\et$) for a system with a planet marked at the circle. We used the GJ\,785 system as reference.
Solid lines represent the collision lines in the case $\Delta \varpi = 0^\circ$ (black) and $\Delta \varpi = 180^\circ$ (blue), while dashed lines delimit the extended crossing orbit regions obtained with $\di$ (Eq.\,\ref{Wisdom1}).
The dashed green lines represent  the ``Hill stability'' limit (Eq.\,\ref{eq.21.gladman}).} 
\label{fig.schema}
 \end{center}
\end{figure}

When the initial orbits are anti-aligned, the extended crossing orbit region is independent on the relation between $\et$ and $\ei$, becoming
{\smallskip}

\begin{center}
\begin{tabular}{c c}     
          Interior limit & Exterior limit  \\
\hline\hline\noalign\\
           $ \at > (\ai - \di) \Frac{1-\ei}{1+\et}$ , & $ \at < (\ai + \di) \Frac{1+\ei}{1-\et} $ \\ 
{\smallskip}\\ \hline
\end{tabular}
\end{center}

In Fig. \ref{fig.schema} we plotted the different limits to be applied. 
In the case of a massless test particle, the orbit of the planet does not precess.
As a consequence, the precession of the test particle orbit will misalign both orbits after some time, so the stability criterion is reduced to the most conservative situation ($\Delta \varpi = 180^\circ$).
However, as we will show in the examples of the following section, when we consider a Earth-like mass planet, the orbits of both planets are coupled, and they may precess together, or present oscillations around an equilibrium configuration.

\section{Testing the stability criterion in real systems} 

\begin{figure*}
 \begin{center}
   \begin{tabular}{c c}
\includegraphics*[width=8.5cm]{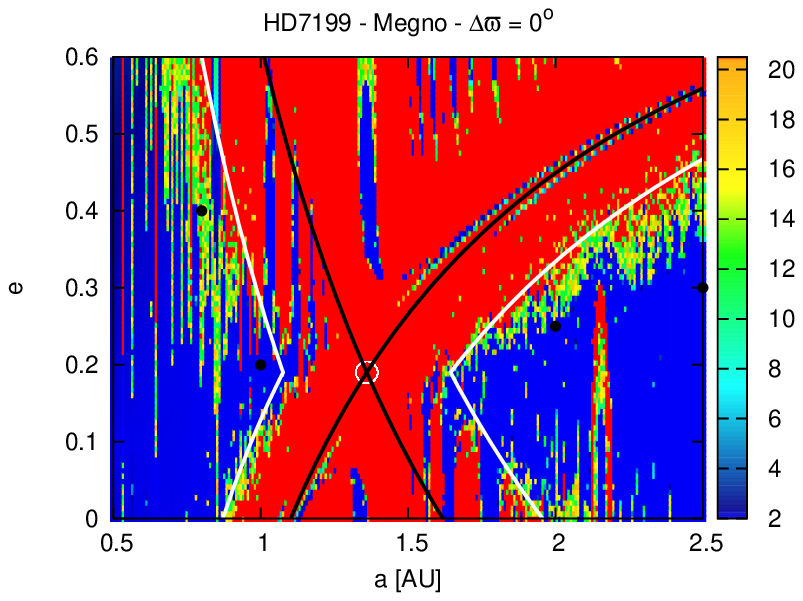} &
\includegraphics*[width=8.5cm]{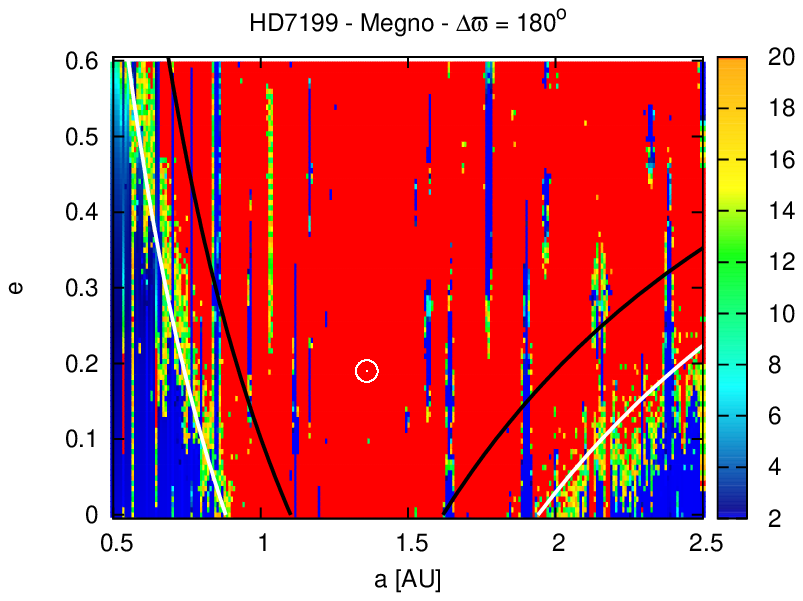}  \\
\includegraphics*[width=8.5cm]{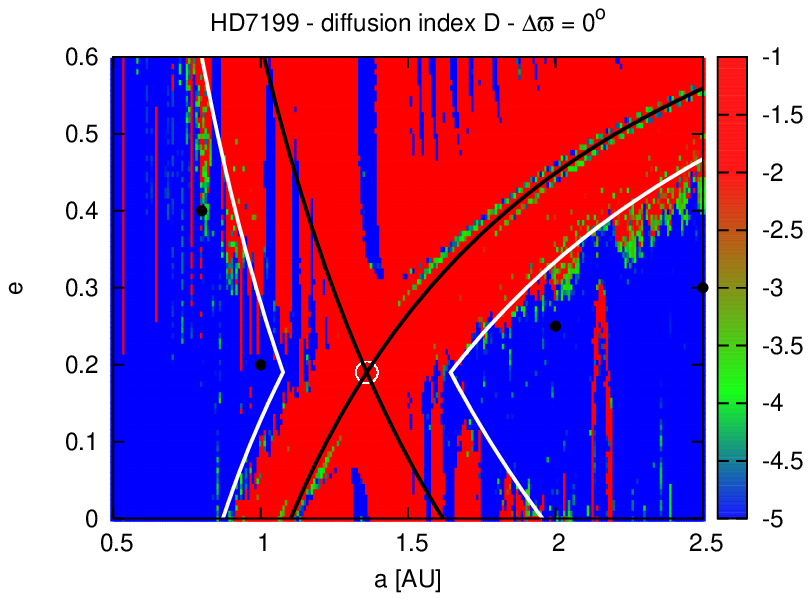} &
\includegraphics*[width=8.5cm]{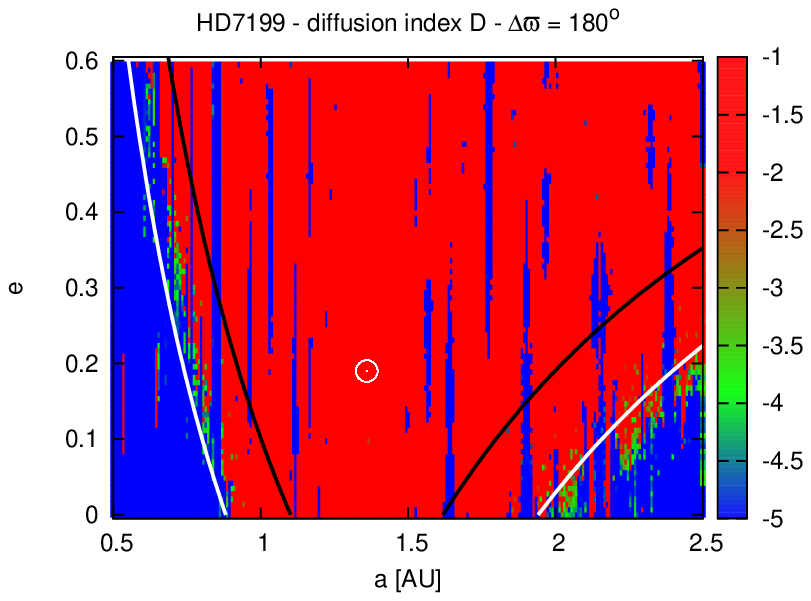} \\                                  
\includegraphics*[width=8.5cm]{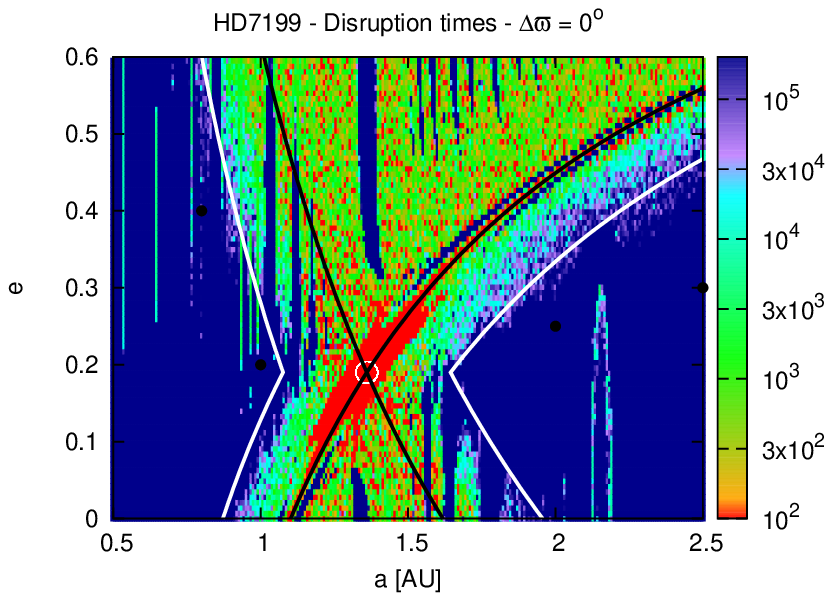} &
\includegraphics*[width=8.5cm]{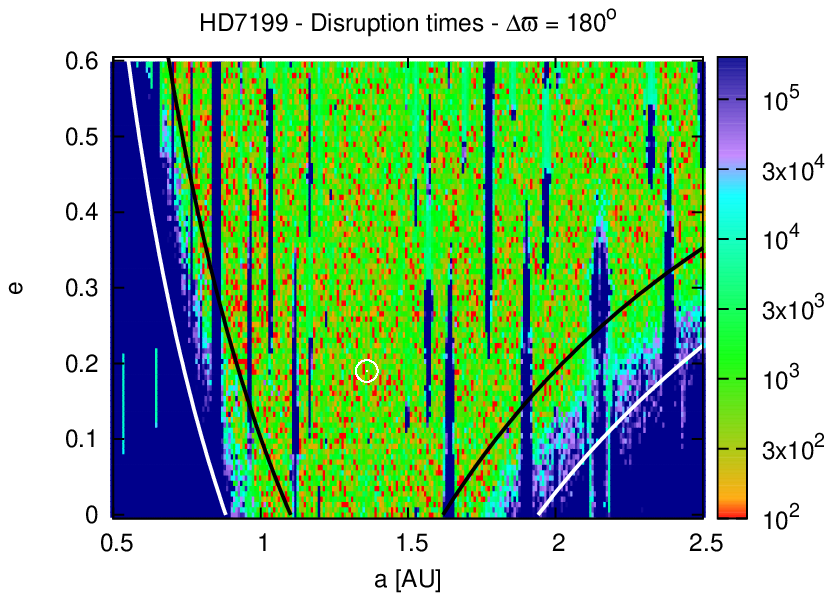} 
   \end{tabular}
 \caption{Stability maps using different chaos indicators for the HD\,7199 planetary system with the inclusion of an additional $2\,M_\oplus$ mass test planet. Top: MEGNO chaos indicator; Middle: Mean motion diffusion using frequency analysis; Bottom: Direct numerical simulations over 200\,kyr. For a fixed initial condition of the observed planet $b$ at $\ai = 1.36$\,AU and $\ei = 0.19$, the phase space of the system is explored by varying the semi-major axis $\at$ and eccentricity $\et$ of the test planet, respectively. The step size is 0.008\,AU in semi-major axis and 0.007 in eccentricity. Left: $\Delta \varpi = 0$; Right: $\Delta \varpi = 180^\circ$. 
 The black dots give the initial conditions for the orbits shown in Figure\,\ref{fig.simdw1}.
} 
\label{fig.hd7199}
 \end{center}
\end{figure*}

\begin{figure*}
 \begin{center}
   \begin{tabular}{c c}
 \includegraphics*[width=8.5cm]{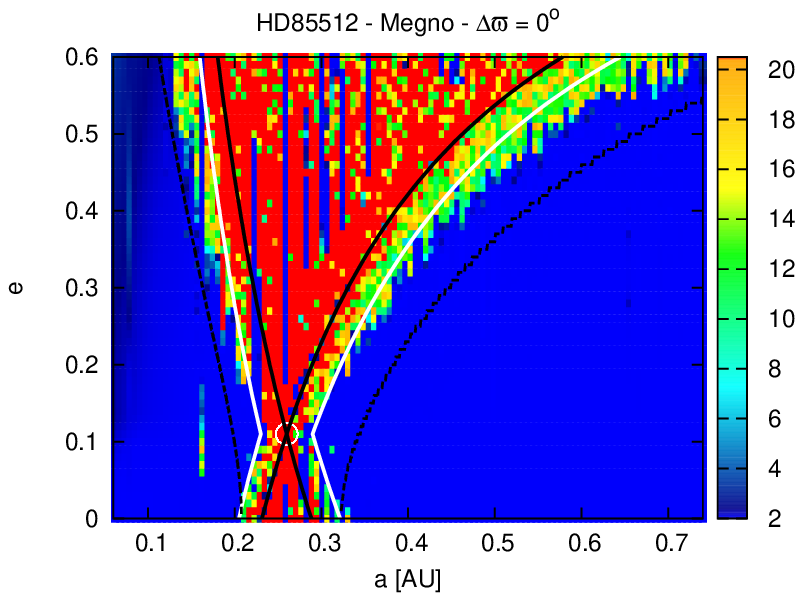} & 
 \includegraphics*[width=8.5cm]{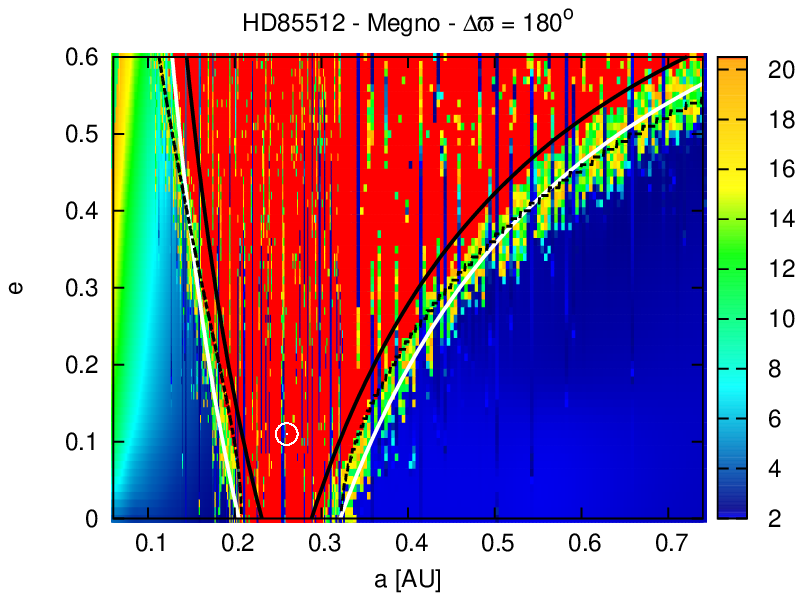}   \\    
 \includegraphics*[width=8.5cm]{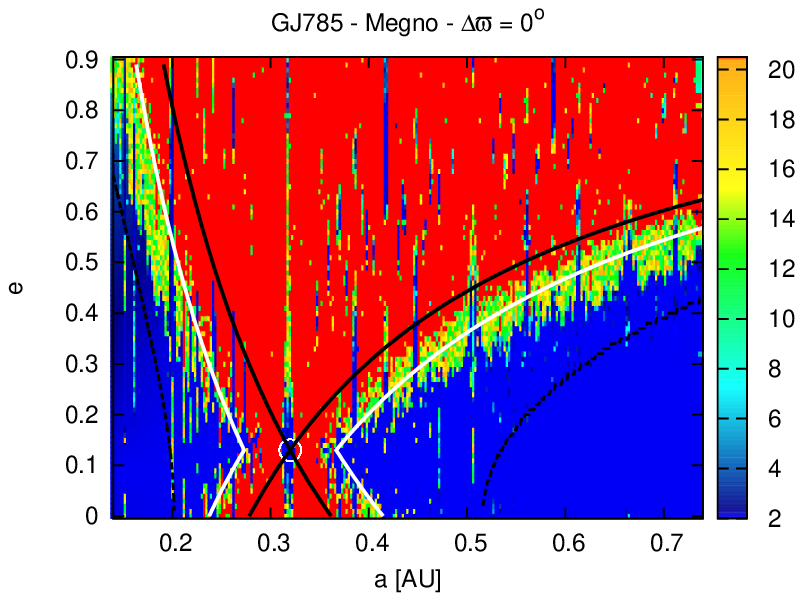}   & 
 \includegraphics*[width=8.5cm]{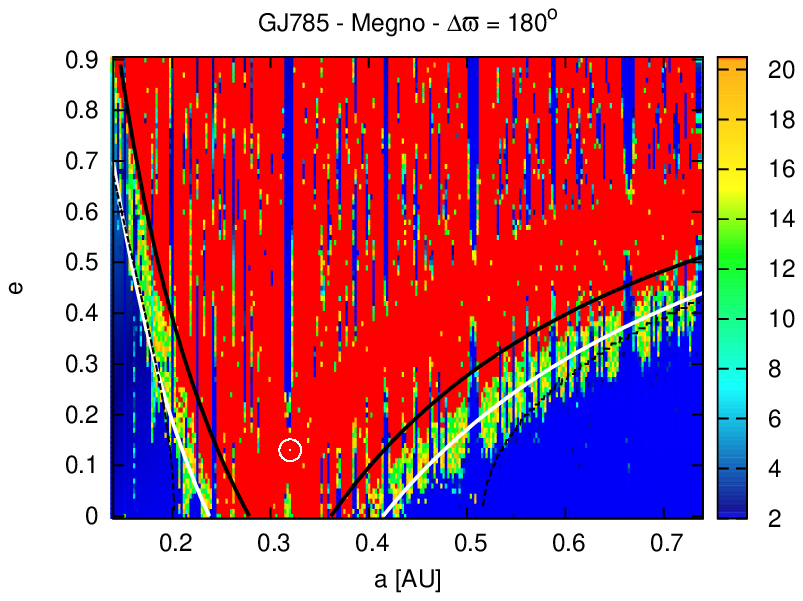}   \\                                                          
\includegraphics*[width=8.5cm]{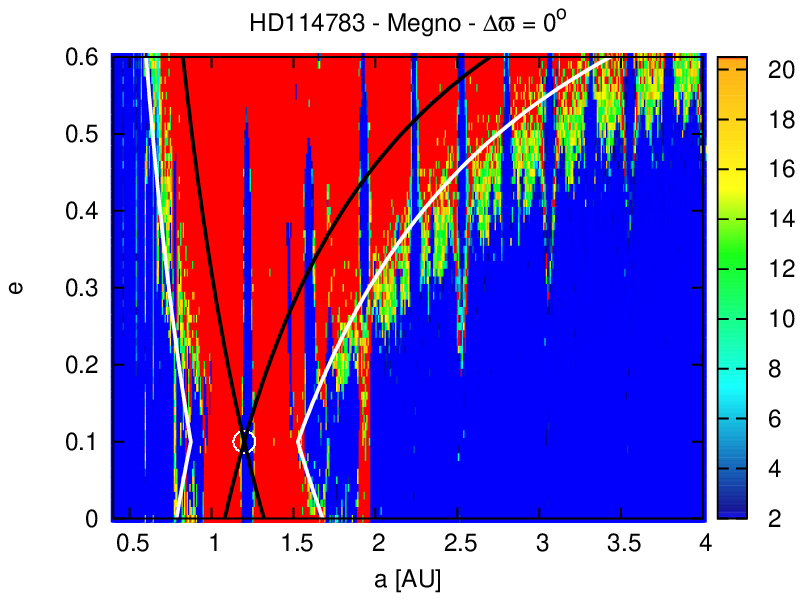} & 
\includegraphics*[width=8.5cm]{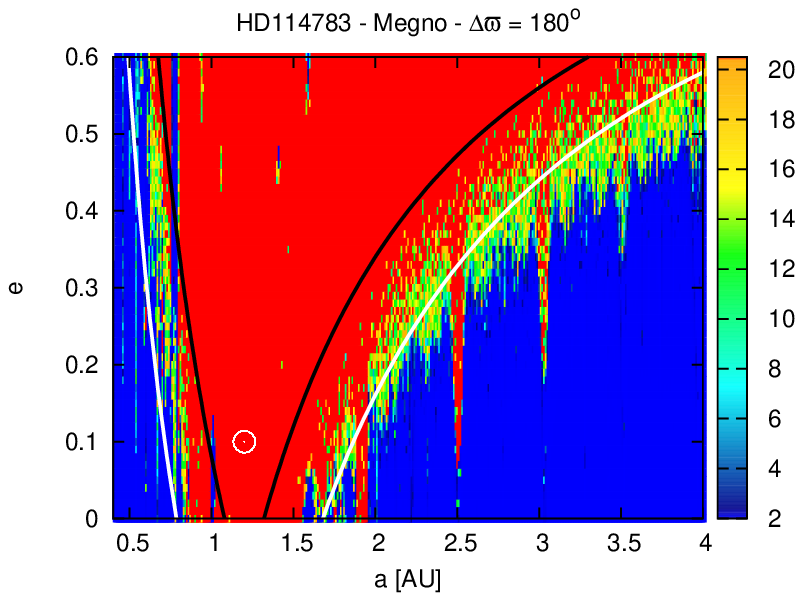}                                     
\end{tabular}
 \caption{Stability maps for an additional test planet $\mt = 2 M_{\oplus}$ for different single-planet systems (with different masses), using the MEGNO chaos indicator. Top: HD\,85512 ($\mi = 3.6 \, M_\oplus$); Middle: GJ\,785 ($\mi = 16.9 \, M_\oplus$); Bottom: HD\,114783 ($\mi = 317.8 \, M_\oplus$). For a fixed initial condition of the observed planet $b$, the phase space of the system is explored by varying the semi-major axis $\at$ and eccentricity $\et$ of the test planet, respectively.  The step size is 0.002\,AU in semi-major axis (0.01\,AU for HD\,114783) and 0.01 in eccentricity. Left: $\Delta \varpi = 0$; Right: $\Delta \varpi = 180^\circ$.
The black dashed line was computed using Eq.\,(\ref{eq.21.gladman}) and gives the limits for ``Hill-stability'' zones.
For identical mass planets there is good agreement between these zones and stability limits defined by the collision lines $\pm \di$. However, for low mass ratios $\mt / \mi$, large stable areas are discarded.} 
\label{fig.one}
 \end{center}
\end{figure*}

We now test the applicability of the crossing orbit criterion for eccentric orbits derived in the previous section by applying it to real systems.
We first select some systems with a single planet to illustrate how the method works, and then  extend the results to systems with two and three planets.

The integrations were made using a Bulirsch-Stoer based N-body code (precision better than $10^{-12}$) using as initial conditions astrocentric  osculating variables. We used averaged MEGNO\footnote{MEGNO is the acronym of Mean Exponential Growth factor of Nearby Orbits.} chaos indicator  $ \langle Y \rangle$ \citep{Cincotta_2000}, to quickly identify the initial conditions leading to chaotic evolution. 
MEGNO is a a wide-spread tool for study the stability of planetary systems, that is able to  distinguish between different levels of chaoticity \citep[for a review see][]{Maffione_etal_2013}.
The total integration time was chosen to be roughly around 50,000 periods of the exterior orbit, which is long enough to put in evidence chaotic behavior.
For comparison, we also computed the stability index ($D$), a diffusion indicator based on frequency analysis \citep[see][]{Laskar_1990,Laskar_1993PD}, and the disruption times for the evolution of the system over 200~kyr.

In Figure\,\ref{fig.hd7199} we compute stability maps for the HD\,7199 planetary system with the inclusion of an additional $2\,M_\oplus$ mass test planet. For a fixed initial condition of the observed planet $b$ at $\ai = 1.36$\,AU and $\ei = 0.19$, the phase space of the system is explored by varying the semi-major axis $\at$ and eccentricity $\et$ of the test planet. We observe that the three methods provide more or less the same results.
The limits in semi-major axis for the grids were chosen to illustrate the validity of the method. The already detected planets are represented by white circles. The black solid lines represent the collision lines, and the white solid lines show  the extended crossing orbit limit as defined in Figure\,\ref{fig.schema}. 
If the system is regular or stable for a given additional planet with mass $\mt$, then an hypothetical lower-mass planet should also be.

\subsection{One planet}

\subsubsection{Stability maps}

\begin{table}
\caption{Initial osculating elements for single-planet systems \citep{vogt_etal_2002, pepe_etal_2011, Dumusque_etal_2011}.}
\label{tab-one-planet} 
\footnotesize
\centering
\begin{tabular}{l|l|c|c|c|}     
 $\#$   & a & e & K & mass  \\
{\smallskip}
 & [AU] &  & [ms$^{-1}$] & [M$_{\oplus}$]   \\
\hline\hline\noalign{\smallskip}
GJ\,785\,b &0.32 &0.13 &3.0 &16.9 \\
HD\,7199\,b &1.36 &0.19 &7.76 &92. \\
HD\,85512\,b &0.26 &0.11 &0.77 &3.6 \\
HD\,114783\,b &1.2 &0.1 &27.0 &317.8 \\
\end{tabular}
\end{table}

We selected four systems to show some examples of how the crossing orbit criterion works, namely GJ\,785, HD\,114783,  HD\,7199, and HD\,85512. GJ\,785\,b is a Neptune-mass planet with minimum mass 16.9 $M_{\oplus}$ with an orbital period of P=74.39 (semi-major axis of 0.32~AU) and eccentricity 0.11, orbiting a K3V dwarf ($m_\star=0.80\,M_\odot$, 
\citealt{pepe_etal_2011}). 
HD\,114783\,b is a Jupiter-like planet orbiting a K2V star ($m_\star=0.92\,M_\odot$) 
with minimum mass 317.8 $M_{\oplus}$, a period of 496.9 days (semi-major axis of 1.2~AU), and eccentricity of 0.1  \citep{vogt_etal_2002}.
HD\,7199 is a K0IV/V star ($m_\star=0.89\,M_\odot$) 
harboring a planet with an orbital period of 615 days (semi-major axis of 1.36~AU), eccentricity 0.19, and a minimum mass of 92~$M_\oplus$ \citep{Dumusque_etal_2011}. HD\,85512 is a K5V star ($m_\star=0.69\,M_\odot$) 
and has a planet with velocity amplitude smaller than one meter per second (0.77 $\pm$ 0.09 ms$^{-1}$), with an orbital period of 58.4 days (semi-major axis of 0.26~AU), eccentricity 0.11, and a minimum mass of 3.6 $M_\oplus$ \citep{pepe_etal_2011}.

The stability analysis considers an hypothetical additional planet with mass equal to 2 $M_{\oplus}$ (or a signal of amplitude of 0.85 ms$^{-1}$ for a planet in a circular orbit at 0.06 AU for a Sun-like star). Blue colors indicate regular orbits, while red colors indicate strongly chaotic orbits that usually lead to collision between planets or escape from the system after close encounters (some of them can collide with the central star). Many orbits with intermediate MEGNO values (yellow colors) were integrated by $10^6$ years ($\sim 6\times 10^6$ orbits of the known planet) and appear to be stable.  The black solid lines represent the collision lines, and the white solid lines show  the extended crossing orbit limit as defined in Figure\,\ref{fig.schema}. The ``Hill-stability'' limits defined by expression (\ref{eq.21.gladman}) are represented by black dashed lines (when this line is not show it  means that it is beyond the semi-major limits of the figure).

Figure~\ref{fig.one} shows the results for the single planet HD\,85512, GJ\,785, and HD\,114783 systems, using different $\Delta \varpi$ as initial conditions. Unstable orbits (red regions) are easily recognized inside the region delimited by the white lines. Regular motion depends strongly on the eccentricity and difference of initial $\Delta \varpi$. We additionally observe some stable regions associated with  MMRs, including  co-orbital configurations, as vertical blue/yellow lines.  The stable regions for initial conditions with $\Delta \varpi=0^\circ$ (left frames) are greater than those with $\Delta \varpi=180^\circ$ (right frames). 

For the HD\,85512 system, which has $\mt / \mi \sim 1$, the ``Hill-stability'' regions defined by expression (\ref{eq.21.gladman}) almost coincide with the regions given by $\di$ with $\Delta \varpi = 180^\circ$ (Fig.\,\ref{fig.one}).
However, for smaller mass ratios $\mt / \mi $, the ``Hill-stability'' limit excludes large regions that can also be considered stable. Indeed, for the GJ\,785 system ($\mt / \mi \sim 0.1$), this limit still gives a good estimation for very eccentric orbits, but becomes a bad approximation when the test planet is near a circular orbit.
For the HD\,114783 system  ($\mt / \mi \sim 0.01$) the  ``Hill-stability'' region starts at around $ \ai = 6$~AU (for $\ei=0$), well beyond the limits shown in Figure~\ref{fig.one}, meaning that a wide region from 2 to 6\,AU  is incorrectly discarded.
Therefore, we see that the ``Hill-stability''  limit is only appropriate for systems with planets of identical mass.  { This is expected since \citet{Marchal&Bozis} criterion cannot be applicable in the restricted problem.}


\begin{figure}
 \begin{center}
   \begin{tabular}{c}
\includegraphics*[width=8.5cm]{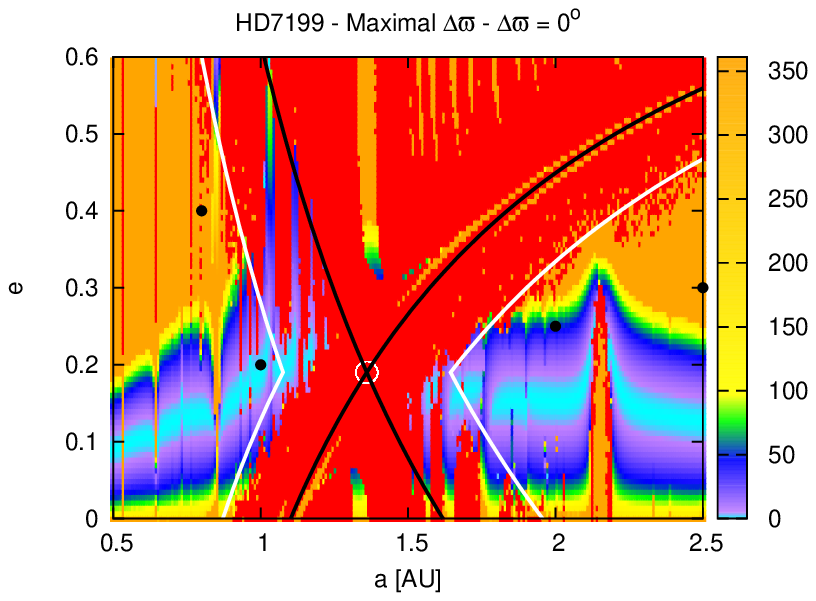} \\
\includegraphics*[width=8.5cm]{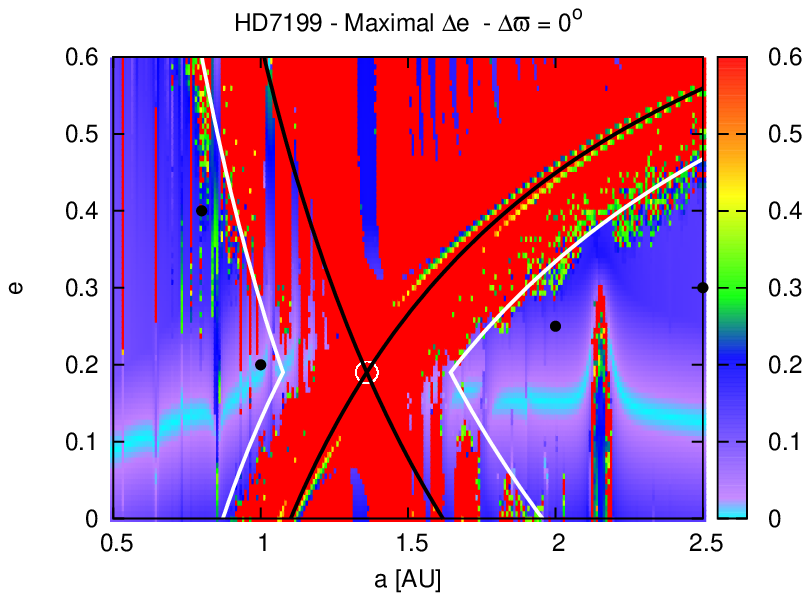} 
   \end{tabular}
 \caption{Maximal libration amplitude for $\Delta \varpi$ (Top), and the maximal amplitude of eccentricity oscillations $\Delta \et$ (Bottom) for an additional $2 \, M_\oplus$ test planet in the  HD\,7199 system. The grid in ($\at$, $\et$) is the same used in Figure\,\ref{fig.hd7199} with $\Delta \varpi = 0$. We observe two different dynamical regimes for $\Delta \varpi $, in particular there is libration of this angle around $0^\circ$ in a strip roughly comprised between $ 0.05 < \et < 0.25 $. We also observe that $\Delta \et$ reproduces quite well the results given by the stability indicators (Fig.\,\ref{fig.hd7199}). The black dots give the initial conditions for the orbits shown in Figure\,\ref{fig.simdw1}.} 
\label{fig.eccdw}
 \end{center}
\end{figure}

\subsubsection{The importance of the initial $\Delta \varpi$}

\label{tiiw}

In section~\ref{coc} we saw that for identical eccentricities, initially aligned orbits ($\Delta \varpi = 0^\circ$) allow closer semi-major axis (Fig.\,\ref{fig.schema}).
If the two orbits precess independently, the angle $\Delta \varpi$ can assume any value and always reach $180^\circ$ after some time.
Therefore, it is often assumed that the initial value of $\Delta \varpi$ is irrelevant, and that the crossing orbit criterion should rely on the most conservative anti-aligned configuration \citep[e.g.][]{Barnes_2006,Tuomi_etal_2013}.
However, in Figures \ref{fig.hd7199} and \ref{fig.one} we can clearly observe that there is a significant difference between the two situations.
To understand why a given configuration is stable for some initial conditions with  $\Delta \varpi = 0^\circ$, but becomes unstable when $\Delta \varpi = 180^\circ$, in Figure~\ref{fig.eccdw} we plot the maximal libration amplitude for $\Delta \varpi$ and the maximal amplitude of eccentricity oscillations of the test planet, $\Delta \et$, for HD\,7199 when $\Delta \varpi = 0^\circ$.

\begin{figure*}
 \begin{center}
   \begin{tabular}{c c c c}
\includegraphics*[width=4.05cm]{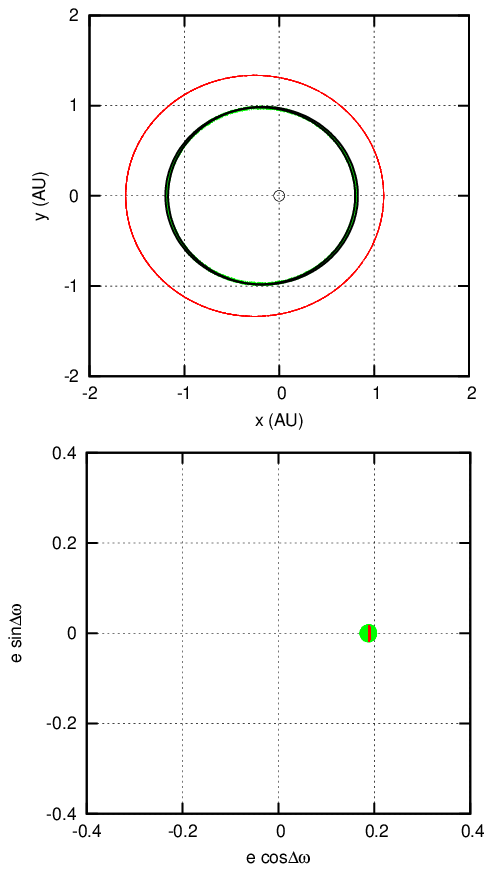} &
\includegraphics*[width=4.05cm]{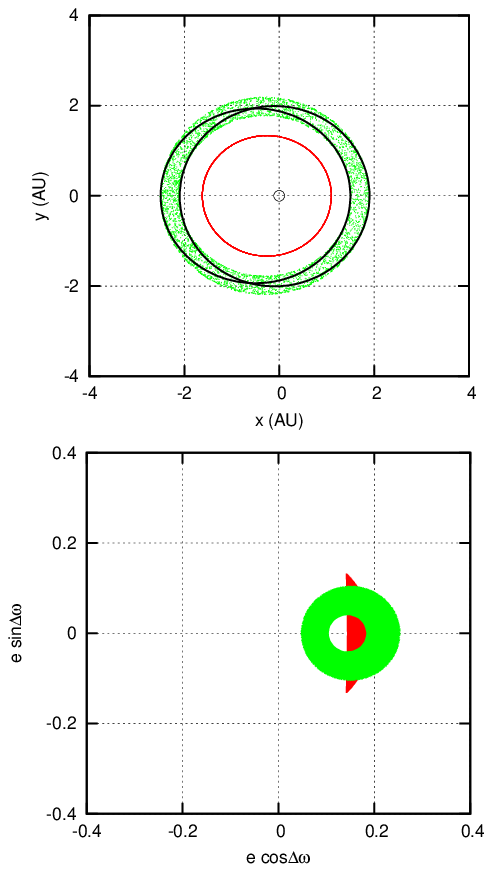} &
\includegraphics*[width=4.05cm]{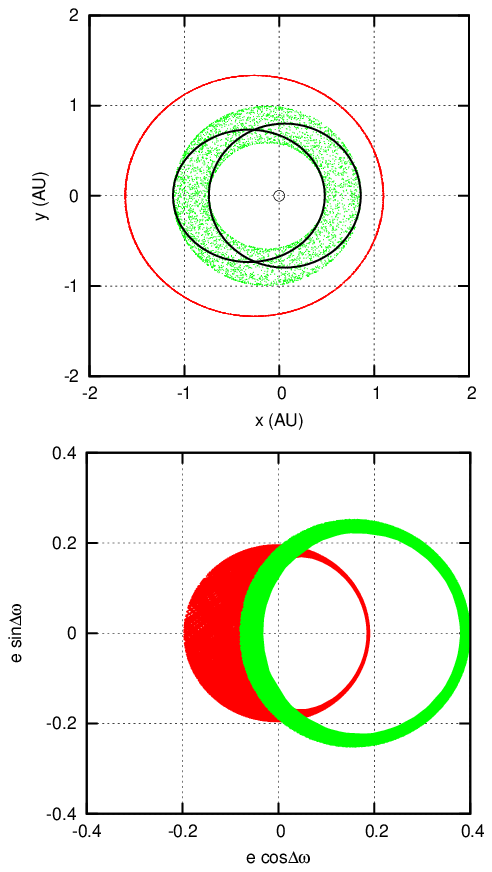} &
\includegraphics*[width=4.05cm]{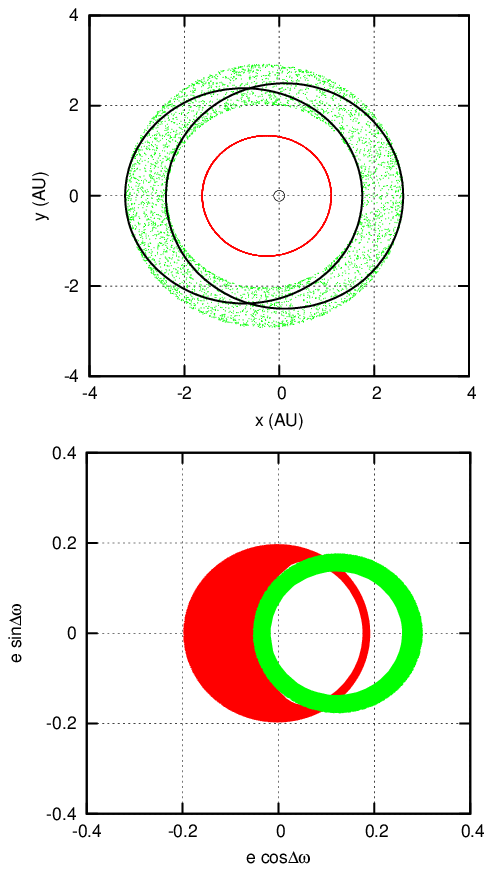} 
   \end{tabular}
 \caption{Long-term evolution of the HD\,7199 planetary system over 1\,Gyr, with the inclusion of an additional $2\,M_\oplus$ test planet, located at initial orbits selected from Figure\,\ref{fig.eccdw}: (a) $\at = 1.0$\,AU, $\et = 0.20$; (b) $\at = 2.0$\,AU, $\et = 0.25$; (c) $\at = 0.8$\,AU, $\et = 0.40$; (d) $\at = 2.5$\,AU, $\et = 0.30$. Each dot corresponds to the position of the planets every 10\,kyr. In the top panels we show a face-on view of the system. $x$ and $y$ are spatial coordinates centered on the star in a frame co-precessing with the observed planet (in red). Solid lines show the orbits with maximal and minimal eccentricity for the test planet (in green). In the bottom panels we show both orbits on the $e \, \mathrm{exp} (i \Delta \varpi)$ plane. In plots (a) and (b) $\Delta \varpi$ is in libration around zero, which prevents close encounters, while in plots (c) and (d) $\Delta \varpi$ circulates, but close encounters also do not occur because for $\Delta \varpi = 180^\circ$ the eccentricity of the test planet is always minimal. } 
\label{fig.simdw1}
 \end{center}
\end{figure*}

We observe that there is a large strip roughly comprised between $ 0.05 < \et < 0.25 $ for which the amplitude of $\Delta \varpi$ is always smaller than $50^\circ$, and it goes down to almost zero for initial conditions close to $\et = \ei$. For all those trajectories we conclude that $\Delta \varpi$ is in libration around $0^\circ$, and therefore overlap between the two orbits is prevented, so they can remain stable (see some examples in Fig.\,\ref{fig.simdw1}a,b).

However, for initial eccentricities outside the  range $ 0.05 < \et < 0.25 $ the angle $\Delta \varpi$ is in circulation, so the orbits become anti-aligned at some point of the evolution. We may then wonder why these trajectories remain stable, while the exact same initial values for ($\at$, $\et$) starting with $\Delta \varpi = 180^\circ$ are not  (Fig.\,\ref{fig.hd7199}).
The reason for this is more subtle, but easily understandable if we plot the eccentricity evolution of those orbits together with $\Delta \varpi$ (Fig.\,\ref{fig.simdw1}c,d).
Since the eccentricities of the two planets are coupled, due to the conservation of the angular momentum, we can write to the first order in eccentricity \citep[e.g.][]{Laskar_etal_2012}:
\begin{equation}
\et^2 = \langle \et^2 \rangle + \Delta \et \cos \Delta \varpi \ , \label{eq.lap}
\end{equation}
where $\langle \et^2 \rangle$ and $\Delta \et$ are constant values.
For orbits starting anti-aligned, the initial value of $\et$ corresponds to the minimal eccentricity, because $\cos \Delta \varpi = -1$. However, for orbits initially aligned, the initial value of $\et$ is already the maximal eccentricity that the planet can attain ($\cos \Delta \varpi = 1$).
Therefore, when an initial aligned orbit becomes anti-aligned, its eccentricity is given by 
$ \et^2 = \et_{0}^2 - 2 \Delta \et $, where $\et_{0}$ is the initial value of $\et$.
Thus, $\et$  is always smaller than the initial higher $\et_{0}$, meaning that the orbit is more circular when becomes anti-aligned and there is no overlap with the other orbit.

{ Due to the conservation of the angular momentum, when $\et$ decreases, $\ei$ increases}.
Therefore, this protecting mechanism is particularly effective when the mass of the test planet is much smaller than the mass of the existing planet, since the eccentricity of the more massive body almost does not change (Fig.\,\ref{fig.one}).
We then conclude that when searching for stability regions around planets in elliptical orbits, the crossing orbit criterion with $\Delta \varpi = 0^\circ$ provides a more general picture for the stability of the system, less conservative than the traditional option with $\Delta \varpi = 180^\circ$. 

Comparing Figures~\ref{fig.hd7199} and \ref{fig.eccdw} we also observe that the maximal amplitude of the eccentricity variations in the stable areas is about 0.2, while unstable areas correspond to larger amplitudes. Thus, it can also be used as a stability indicator, as proposed by \citet{Giuppone_etal_2012}. 
Although this indicator is less sensitive to some subtle dynamical regimes, it can provide a global picture much faster than the traditional stability indicators shown in Figure\,\ref{fig.hd7199}.

\subsection{Multi-planet systems}

In order to test the validity of the stability criterion presented in section~\ref{coc} for systems hosting more than one planet, { we also tested it in several existing multi-planet systems.
Here we present the results for} HD\,47186, HD\,51608 (two planets, Table\,\ref{tab-hd47186}), and HD\,134606 (three planets, Table\,\ref{tab-hd134606}).

\begin{figure*}
 \begin{center}
   \begin{tabular}{c c}
\includegraphics*[width=8.5cm]{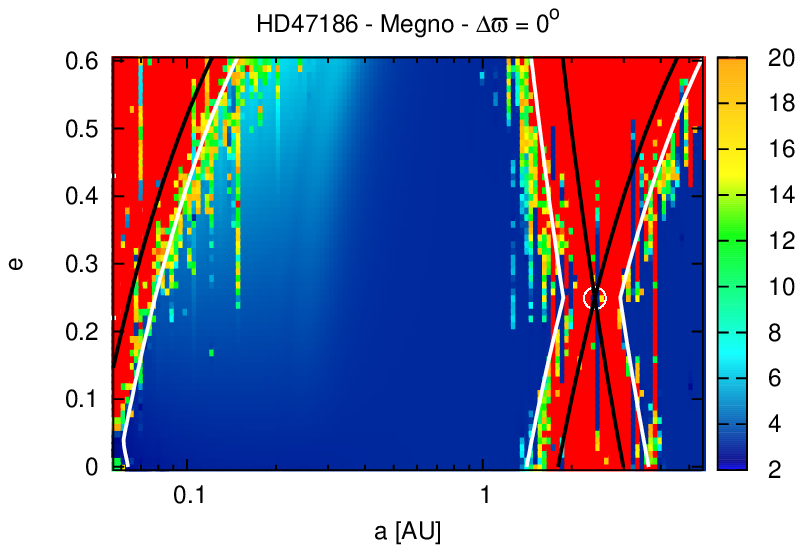} &
\includegraphics*[width=8.5cm]{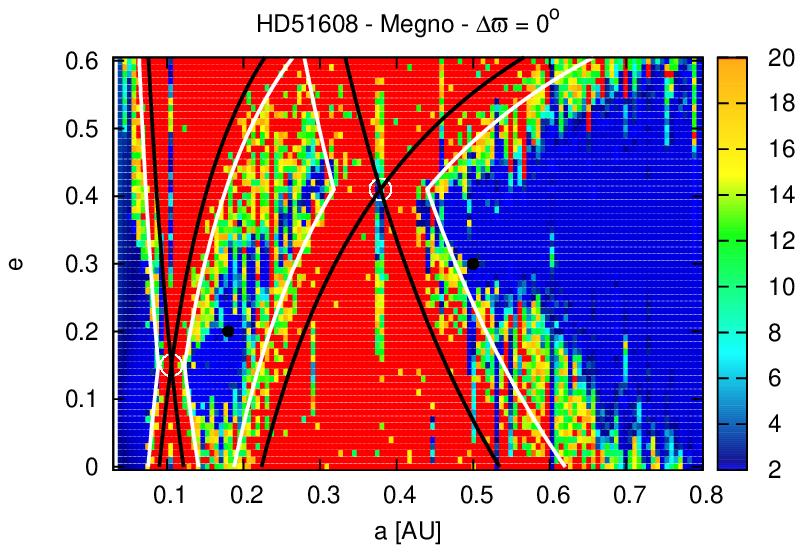} \\
\includegraphics*[width=8.5cm]{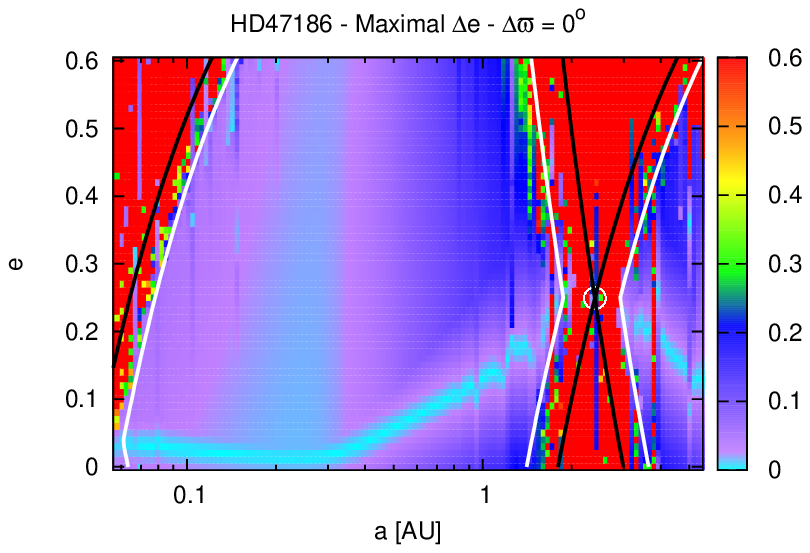} &
\includegraphics*[width=8.5cm]{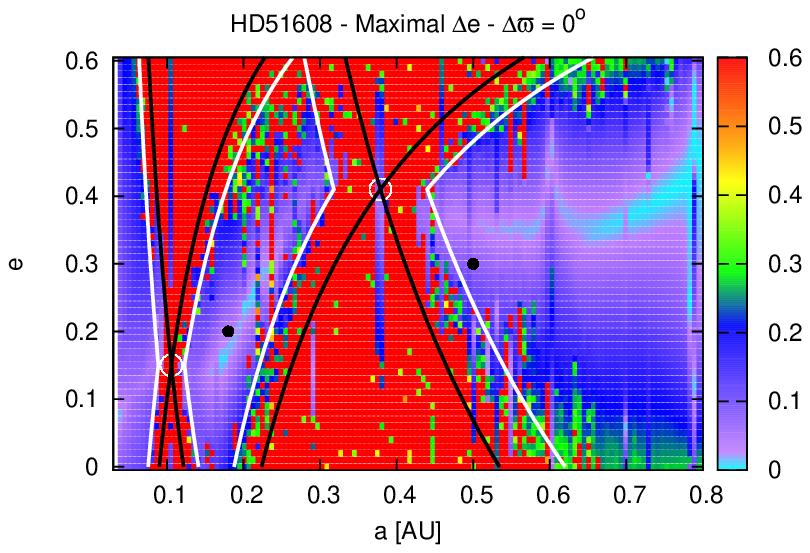}
   \end{tabular}
 \caption{Stability maps for an additional third test planet $\mt = 2 \, M_{\oplus}$ for different two-planet systems (with different semi-major axis ratios), using the MEGNO chaos indicator over 200 kyr (Top), and the eccentricity amplitude of test planet (Bottom). Left: HD\,47186 ($a_2/a_1 = 47.9$); Right: HD\,51608 ($a_2/a_1 = 3.6$). For a fixed initial condition of the two observed planets $b$ and $c$ (Table\,\ref{tab-hd47186}), the phase space of the system is explored by varying the semi-major axis $\at$ and eccentricity $\et$ of the test planet, respectively. 
 The step size is 0.05 in a log-scale of semi-major axis (left) or 0.002~AU (right), and 0.01 in eccentricity.{ The black dots give the initial conditions for the orbits shown in Figure\,\ref{fig.stab2}}.}
\label{fig.hd47186}
 \end{center}
\end{figure*}

The G5V star HD\,47186 ($m_\star=0.99\,M_\odot$) 
harbors two Neptune-like planetary companions in hierarchical orbits as reported by \citet{Bouchy_etal_2009}, with orbital periods of 4 and 1353 days.
On the other hand, HD\,51608 is a K0IV star ($m_\star=0.80\,M_\odot$) 
harboring two super-earth planetary companions, but in a more compact configuration, with orbital periods of 14 and 95 days \citep{Mayor_etal_2011}.
We performed numerical simulations using the initial osculating elements given in Table \ref{tab-hd47186}.
In Figure \ref{fig.hd47186}
 we show the MEGNO stability map for an additional $2 \, M_\oplus$ test planet.

As for previous stability maps for single-planet systems (Figs.\,\ref{fig.hd7199} and \ref{fig.one}), around each planet the stability regions appear to be relatively well delimited by the collision pericentric and apocentric lines $\pm \di$ (section~\ref{coc}).
For the HD\,47186 planetary system this result was expected, since the gravitational interactions between the two observed planets are weak 
due to the large mutual distances between the planets.
Indeed,  when the innermost planet is not considered, no significative difference is found in the stability map of the more massive planet.

For the HD\,51608 planetary system the two orbits are relatively close, so we can expect to observe three-body resonances with the test planet. However, this additional source of instability does not appear to be very strong:
stable and regular motion is still possible between the already discovered planets, where
 the global stability regions { continue to be reasonably} well described by the extended crossing orbit criterion (described in section\,\ref{coc}).
{ In order to test the reliability of the MEGNO stability indicator in a such extreme situations, we selected two sets of initial conditions (one inside and another outside the orbit of the outermost planet) and performed a direct numerical integration of the system over 100~Myr (Fig.~\ref{fig.stab2}).
We observe that the system remains stable.}
In Figure\,\ref{fig.hd47186}, together with the MEGNO stability indicator, we also plot the amplitude of the eccentricity variations of the test planet, $\Delta \et$. 
As for the single-planet system HD\,7199, we observe that $\Delta \et$ is also still able to reproduce the global stability in the case of a multi-planet system, although it is less accurate in predicting the secular chaos.

\begin{table}
\caption{Initial osculating elements for HD\,47186 \citep{Bouchy_etal_2009} and HD\,51608 \citep{Mayor_etal_2011} planetary systems.}
\label{tab-hd47186} 
\footnotesize
\centering
\begin{tabular}{l|l|c|c|c|}     
 $\#$   & a & e & K & mass  \\
{\smallskip}
 & [AU] &  & [ms$^{-1}$] & [M$_{\oplus}$]   \\
\hline\hline\noalign{\smallskip}
HD\,47186\,b &0.05 &0.038 &9.12 &22.78 \\
HD\,47186\,c &2.395 &0.249 &6.65 &111.42 \\
\noalign{\smallskip}\hline\noalign{\smallskip}
HD\,51608\,b &0.1059 &0.15 &4.10 &13.14 \\
HD\,51608\,c &0.379 &0.41 &3.25 &17.97 
\end{tabular}
\end{table}



\begin{table}
\caption{Initial osculating elements for the HD\,134606 planetary system \citep{Mayor_etal_2011}.}
\label{tab-hd134606} 
\footnotesize
\centering
\begin{tabular}{l|l|c|c|c|}     
 $\#$   & a & e & K & mass  \\
{\smallskip}
 & [AU] &  & [ms$^{-1}$] & [M$_{\oplus}$]   \\
\hline\hline\noalign{\smallskip}
HD\,134606\,b &0.102 &0.15 &2.68 &9.27 \\
HD\,134606\,c &0.296 &0.29 &2.17 &12.14 \\
HD\,134606\,d &1.157 &0.46 &3.66 &38.52
\end{tabular}
\end{table}

\begin{figure}
 \begin{center}
   \begin{tabular}{c}
\includegraphics*[width=8.5 cm]{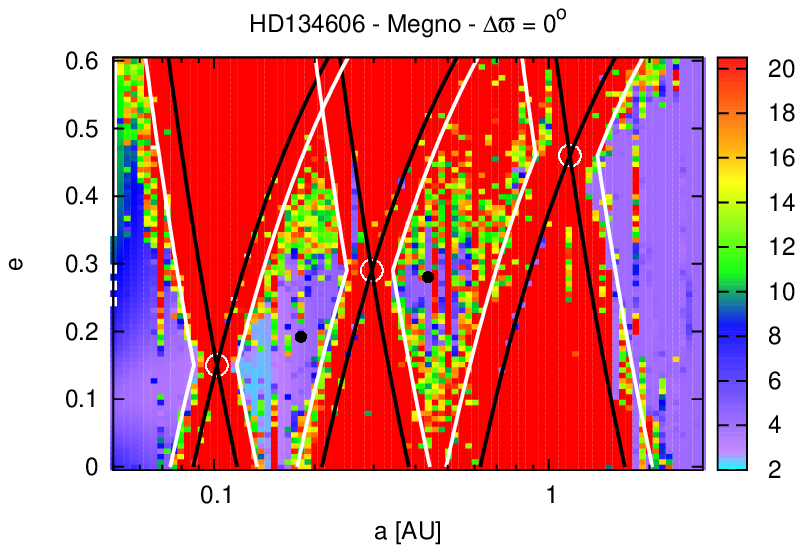}  \\  
\includegraphics*[width=8.5 cm]{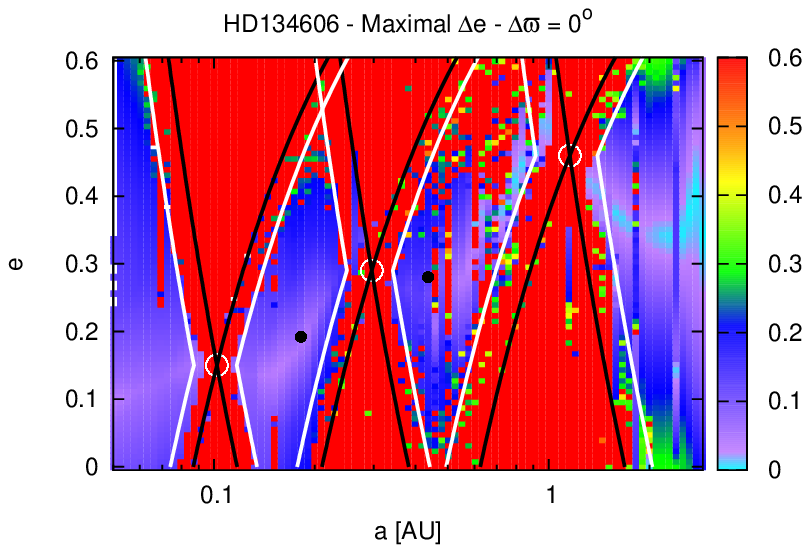}                                
   \end{tabular}
 \caption{Stability maps for an additional fourth test planet $\mt = 2\,M_{\oplus}$ for the HD\,134606 three-planetsystem, using the MEGNO chaos indicator over 50 kyr (Top), and the eccentricity amplitude of the test planet (Bottom). For a fixed initial condition of the three observed planets $b$, $c$ and $d$ (Table\,\ref{tab-hd134606}), the phase space of the system is explored by varying the semi-major axis $\at$ and eccentricity $\et$ of the test planet, respectively. The step size is 0.05 in a log-scale of semi-major axis (AU) and 0.008 in eccentricity. { The black dots give the initial conditions for the orbits shown in Figure\,\ref{fig.stab2}}.}
\label{fig.hd134606-megr}
 \end{center}
\end{figure}

\begin{figure}
 \begin{center}
\includegraphics*[width=8cm]{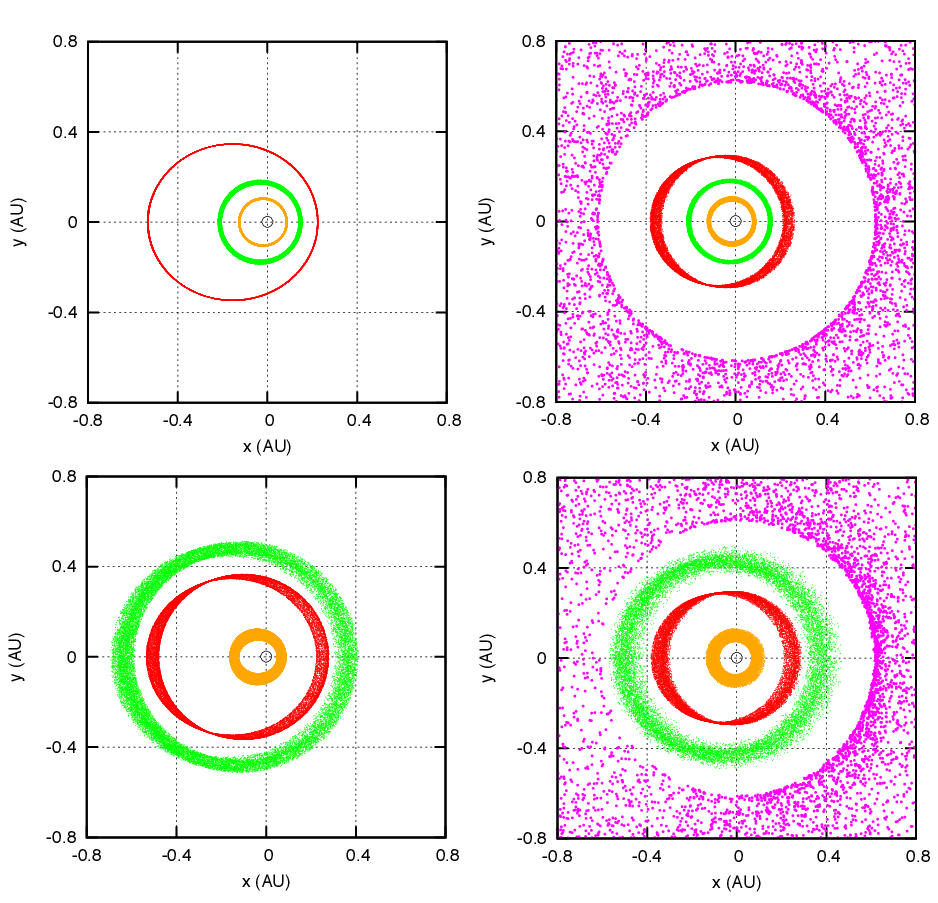} 
 \caption{ Long-term evolution of the HD\,51608 (left) and HD\,134606 (right) planetary systems over 100\,Myr, with the inclusion of an additional $2\,M_\oplus$ test planet (in green), located at initial orbits selected from Figures\,\ref{fig.hd47186} and \ref{fig.hd134606-megr}, repectively: $\at = 0.18$\,AU, $\et = 0.20$ (top), $\at = 0.50$\,AU, $\et = 0.30$ (bottom) for HD\,51608 (left), and $\at = 0.182$\,AU, $\et = 0.192$ (top), $\at = 0.436$\,AU, $\et = 0.280$ (bottom) for HD\,134606. We show a face-on view of the system, each dot corresponding to the position of the planets every 10\,kyr. $x$ and $y$ are spatial coordinates centered on the star in a frame co-precessing with planet \#2 (in red). } 
\label{fig.stab2}
 \end{center}
\end{figure}

HD\,134606 is a G6IV star ($m_\star=0.98\,M_\odot$) 
harboring three Super-Earth planets with orbital periods 12, 60, and 460 days, and moderate eccentricities \citep{Mayor_etal_2011}.
In Figure\,\ref{fig.hd134606-megr} we show 
the stability maps using MEGNO and the amplitude of the eccentricity variations $\Delta e$ in the plane ($a, e$) for an additional fourth-planet with $2\,M_{\oplus}$, with osculating initial elements from Table\,\ref{tab-hd134606}.
Neither the arguments of the pericenter nor the mean anomalies were published in \citet{Mayor_etal_2011}.
However, since the system does not present any mean motion resonance, the mean anomaly is not a critical parameter.
The argument of the pericenter is only important if the system is locked in some kind of equilibrium as the one described in section~\ref{tiiw}.
We assumed for simplicity that all initial $\omega_i = 0$, which corresponds to the most favorable configuration for the stability of an additional fourth planet.
Since the HD\,134606 system contains three planets,
we also computed $\delta$ for the superposition of three body resonances (Eq.\,\ref{Quillen11}).  
However, due to the low mass of the planets, this value almost coincide with $\di$ given by expression (\ref{Wisdom1}), and the differences are not perceptible (see Fig.\,\ref{fig-delta}).
In Figure\,\ref{fig.hd134606-megr} we see that even for a three-planet system of identical masses, the crossing orbit criterion (section~\ref{coc}) still  delimits { reasonably} well the stable regions. { Nevertheless, in this case we can observe several mean motion resonances that destabilize a significant zone inside the ``predicted'' stable regions.
Therefore, we must be conservative when applying the crossing orbit criteria to closely packed multi-planet systems, stable regions for high eccentricity can only be confirmed by using a stability map analysis or by performing long-term numerical simulations
(Fig.~\ref{fig.stab2}).}


\section{Conclusions}

We proposed a {\bf semi-empirical} crossing orbit stability criterion for eccentric planetary systems which is based on Wisdom's criterion of first order mean motion resonance overlap (section~\ref{coc}). In order to test the validity of this criterion we integrated the equations of motion for several single planet systems 
considering an additional low mass planet. We obtained the regions of stability in the plane $(a,e)$ and we observed that this criterion works very well for these systems.

{\citet{Deck_etal_2013} recently obtained an analytic criterion for the overlap of first order mean motion resonances valid in 2 planet systems which has the same functional form as Wisdom's criterion but the mass ratio includes  both planets' masses. As discussed in Sect.~2 our semi-empirical extension of Wisdom's criterion is larger, at most, by a factor 1.76 than  \citet{Deck_etal_2013} analytic criterion. This means that, in practice, we obtain a wider, more conservative estimate for the extent of the chaotic region.   \citet{Mustill_Wyatt_2012} addressed the problem of stability in the circular restricted three-body when the test particles have eccentric orbits, and deduced a  new scaling law for the overlap of first order mean motion resonances which differs from Wisdom's result. However, this criterion was only tested for eccentricities $e\le 0.1$ and, as shown recently by \citet{Deck_etal_2013} it does not explain all the observed chaos even in the low eccentricity regime. Here, we tested a semi-empirical criterion in real planetary systems with high eccentricities which can go up to $e=0.6$ hence a direct comparison with the work of \citet{Mustill_Wyatt_2012} is not possible at this time. }

We tested the crossing orbit criterion on multi-planetary systems with masses $\mi=10^{-6}\,M_\odot$ to $\mi \sim 10^{-3}\,M_\odot$. In this mass range, the chaotic region due to the overlap of 3-body resonances \citep{Quillen2011} is close to the chaotic region estimated from the crossing orbit criterion (Eq.\,\ref{Wisdom1}). Therefore, our results seem to be still approximately valid for multiple planet systems, as exemplified in the cases of 2-planet systems HD\,47186 and HD\,51608, and  3-planet system HD\,134606.
However, when applying the crossing orbit criterion to closely packed multi-planet systems, stable regions for high eccentricity can only be confirmed by using a stability map analysis or by performing long-term numerical simulations.

We additionally explored another 30 non-resonant existing exoplanet systems (but not showed in this paper) and our conclusion remains valid for all of them.
Thus, the crossing orbit criterion seems to provide a quick tool to infer the stability of potential  additional planets in existing extra-solar systems. 

We also showed  that in planetary systems with elliptical orbits, the crossing orbits criterion with $\Delta \varpi = 0^\circ$ provides a more general picture of the stability of the system, less conservative than the traditional option with $\Delta \varpi = 180^\circ$. Finally, we saw that the eccentricity variation  ($\Delta e$) of the test planet during the integration can be used as complement to other chaos indicators, that has the advantage of being computed faster.

\subsection*{Acknowledgements}
We acknowledge financial support from FCT-Portugal (grant PEst-C/CTM/LA0025/2011). The computations were performed on the Blafis cluster at the University of Aveiro.

\bibliographystyle{mn2e}

\bibliography{library}

\appendix

\end{document}